\documentclass[a4paper,12pt]{article}

\usepackage{graphics,psfrag}
\usepackage{amsmath,amsthm,amssymb,epsfig,euscript,array,cite,cancel,color}
\usepackage{cases,empheq}
\usepackage[colorlinks, linkcolor=black, citecolor=blue]{hyperref}
\usepackage{verbatim}
\usepackage{ulem}

\usepackage[paper=a4paper,dvips,top=2.54cm,left=2.2cm,right=2.2cm,
    foot=1cm,bottom=2.5cm]{geometry}

\newcommand{\be}{\begin{equation}}
\newcommand{\ee}{\end{equation}}
\newcommand{\bit}{\begin{itemize}}  
\newcommand{\eit}{\end{itemize}}
\def\bea{\begin{eqnarray}}
\def\eea{\end{eqnarray}}

\def\nn{\nonumber}

\def\a{\alpha}      
     
      \def\G{\Gamma}  
\def\d{\delta}            
\def\e{\epsilon}         
\def\f{\phi}

\def\l{{\lambda}}      \def\L{\Lambda}  
\def\m{\mu}  
\def\n{\nu}  
\def\om{\omega}     \def\Om{\Omega}  
\def\p{\pi}                  
  
         \def\S{\Sigma}  
\def\t{\tau}  
\def\th{\theta}          \def\Th{\Theta}                   
  
\def\z{\zeta}  

\def\na{\nabla}

 \def\cB{{\cal B}}

  \def\cL{{\cal L}}  
    
  \def\cR{{\cal R}}  
 \def\cT{{\cal T}}

\def\pl{\partial}
\def\w{{\wedge}}

\def\omt{\tilde{\omega}}

\def\tt{\th_{\mbox{\tiny T}}}
\def\tl{\th_{\mbox{\tiny L}}}
\def\Lef{\Lambda_{\mbox{\tiny eff}}}
\def\lef{\ell_{\mbox{\tiny eff}}}
\def\OmH{\Omega_{\mbox{\tiny H}}}
\def\TH{T_{\mbox{\tiny H}} }
\def\kH{\kappa_{\mbox{\tiny H}} }

\def\in{{\mbox i}_\xi}
\def\iz{{\mbox i}_\zeta}
\def\dr{{\mbox d}}
\def\Dr{{\mbox D}}

\def\nd{{\nabla}}
\def\ndt{{\tilde \nabla}}

\title{\bf { \Large Quasi-local Energy in 3D Gravity with Torsion} }

\author{ Cheng-Hao Wei\footnote{weichenghao@stu.scu.edu.cn}$~^a$
and Bo Ning\footnote{ningbo@scu.edu.cn}$~^a$ 
\\
\\
{\small\it $^a$ College of Physics, Sichuan University, Chengdu 610065, China}
}

\date{}

\begin{document}

\maketitle

\abstract{ 

\hspace{2mm}

We show that for generic stationary spacetime and specific Killing fields, Wald's approach for quasi-local energy could be generalized to the first order formalism straightforwardly without introducing the Lorentz-Lie derivative. Via this approach, we derive the general formula for the black hole entropy in three dimensional torsional Mielke-Baekler gravity, reproducing precisely the total energy, the angular momentum as well as the black hole entropy for the BTZ-like solutions.

\thispagestyle{empty}
\newpage

\tableofcontents


 \section{Introduction} \label{sec:intro}
 
\hspace{5mm}
 
For gravitational field, there is no well defined local energy density because of the equivalence principle, since at a point one can always find a local frame of reference in which there is no gravitational field \cite{Misner:1974qy}. The best work one could do alternatively, is to define the so-called quasi-local energy. There are many distinct approaches \cite{Szabados:2004vb}, for example, the Komar integral \cite{Komar:1958wp}, the Brown-York approach which is based on the Hamilton-Jacobi method \cite{Brown:1992br}, Nester {\it{et al\,}}'s approach in covariant Hamiltonian formalism \cite{Chen:1998aw, Chang:1998wj}, and the covariant phase space method developed by Wald {\it{et al}} \cite{Lee:1990nz, Wald:1993nt, Iyer:1994ys, Wald:1999wa}. It was shown via Wald's approach that the black hole entropy is the Noether charge of diffeomorphism invariance evaluated at the horizon. This approach was generalized to diffeomorphism covariant gravitational theories with Chern-Simons term by Tachikawa \cite{Tachikawa:2006sz}.

\hspace{5mm}

It was realized that a straightforward generalization of the Wald's approach to the first order formalism of gravitational  theory leads to apparent puzzle  \cite{Jacobson:2015uqa}. This issue is attributed to the fact that the vielbeins are in general not invariant under the flow of the Killing field. By introducing the Lorentz-Lie derivative \cite{Kosmann:1971, Jackiw:1979ub, Obukhov:2006ge, Fatibene:2011}, Jacobson and Mohd reformulated Wald's approach and concluded that the black hole entropy is the horizon Lorentz-diffeomorphism Noether charge  \cite{Jacobson:2015uqa}. Following this idea, generalizations to Chern-Simons-like theories of gravity \cite{Bergshoeff:2014bia} are developed in \cite{Setare:2015nla, Setare:2015cvv, Setare:2015gss, Setare:2017wuj, Adami:2017phg}. 

\hspace{5mm}

We notice, however, that for generic axially symmetric stationary spacetime, it is natural to choose orthogonal frames that are invariant under the flow of a series of special Killing vectors, which are responsible for the definition of the total energy, the angular momentum as well as the black hole entropy. In this case, Wald's approach could be applied straightforwardly in the first order formalism without introducing the Lorentz-Lie derivative, and the generalization to the  Chern-Simons-like theories are much simpler. As pointed out in \cite{Jacobson:2015uqa}, such vielbeins are singular at the horizon and lead to divergent spin connection on the bifurcation surface, which is crucial for giving rise to non-vanishing black hole entropy via Wald's formalism in the first order formalism. We generalize this treatment to gravitational theory with torsion in three dimension, obtaining a novel and consistent approach for deriving the total energy, the angular momentum as well as the black hole entropy in the first order formalism for three dimensional torsional gravity. 

\hspace{5mm}

We apply this approach to a special gravitational theory, i.e., the Mielke-Baekler (MB) model \cite{Mielke:1991nn,Baekler:1992ab}. This is a topological gravity with torsion in three dimension which could be formulated as a gauge theory. It had been realized that gravity could be formulated as a gauge theory of the Poincar\'e group (rather than the Lorentz group), in which torsion is the gauge field strength associated with translations \cite{MacDowell:1977jt}. Unlike its cousin in four dimension, i.e., the Einstein-Cartan theory, in which fermion matter is necessary for non-vanishing torsion on shell, the MB model has non-trivial vacuum solutions with torsion, i.e., the BTZ-like black holes. We derive the general entropy formula for stationary black holes in the MB model (see eq. (\ref{entropyformula})), and the total energy, the angular momentum as well as black hole entropy for the BTZ-like solutions, which are consistent with the first law of black hole thermodynamics as well as previous results in literature via different methods. 

\hspace{5mm}

The rest of this paper is organized as follows. In Section {\ref{sec:wald}} we first review Wald's approach for quasi-local energy and its generalizations, then show that in the first order formalism, Wald's approach could be applied straightforwardly for stationary spacetime and specific Killing vectors. In Section {\ref{sec:quasi}}, we apply our method to the torsional Mielke-Baekler gravity, calculating the total energy, the angular momentum as well as the black hole entropy for the BTZ-like solutions. We conclude in Section \ref{sec:conclusion} with a brief discussion.

\hspace{2mm}


\section{Wald's approach revisited} \label{sec:wald}

\hspace{2mm}

In this section, we first review Wald's approach \cite{Lee:1990nz, Wald:1993nt, Iyer:1994ys, Wald:1999wa} and its various generalizations, especially in the first order formalism. We then  demonstrate that there is an alternative way to generalize Wald's approach straightforwardly in the first order formalism for a large variety of spacetimes with certain symmetries. 

\hspace{2mm}


\subsection{Wald's approach and beyond} \label{sec:review}

\hspace{5mm}

For a diffeomorphism invariant gravitational theory defined by Lagrangian $n$-form $L$ ($n$ is the dimension of the  spacetime), denoting all the dynamical fields by $\f$\,, the variation of the Lagrangian due to $\d \f$ is given by 
\be 
\d L \,=\, E \,\d \f \,+\, \dr \Th(\f, \d \f) \,, 
\ee
where $E$ defines the equations of motion by $E = 0$. The $(n-1)$-form $\Th$, constructed locally out of $\f$, $\d \f$ and  linear in $\d \f$, is called the ``symplectic potential''. Defining the ``symplectic current'' $(n-1)$-form by anti-symmetrizing the variation of $\Th$\,: 
\be
\Om (\f, \d_1 \f, \d_2 \f) \,=\, \d_1 \Th(\f, \d_2 \f) \,-\, \d_2 \Th(\f, \d_1 \f) \,,
\ee
the integration of $\Om$ over a Cauchy surface $\S$ then defines a (pre-)symplectic form in the space of field configuration, which is identified as the phase space of the theory.

\hspace{5mm}

Consider the variation due to an infinitesimal diffeomorphism generated by a vector field $\xi$ : 
\be  \d _\xi \f  \,=\, {\cal L}_\xi \f \,,\ee
the diffeomorphism invariance of $L$ implies the variation of the Lagrangian is 
\be \label{deltaxiL}
\d_\xi L \,=\, {\cal L}_\xi L \,=\, \dr \,\in L \,, 
\ee
which is a total derivative, showing that $\xi$ generates a symmetry. For each $\xi$ there is an associated Noether current $(n-1)$-form: 
\be
j_\xi \,=\, \Th(\f, {\cal L}_\xi \f) \,-\, \in L \,, \label{NoetherJ}
\ee
one could easily check that
\be
\dr  j_\xi \,=\, -\,E  {\cal L}_\xi \f  \,,
\ee
hence $j_\xi$ is closed on-shell, which implies
\be
j_\xi \,\approx\, \dr Q_\xi \,,
\ee
where $Q_\xi$ is the Noether charge $(n-2)$-form. 

\hspace{5mm}

The Hamiltonian $H_\xi$\,, which generates the phase space flow  corresponding to the diffeomophism generated by $\xi$\,, is related to the symplectic form through Hamilton's equation
\be
\d H_\xi = \int_\S \Om(\f, \d \f, {\cal L}_\xi \f) \,, \label{HamiltonsEq}
\ee
in which $\S$ is a Cauchy surface. The above integration turns out to be a boundary integral on shell,
\bea
\d H_\xi &=&  \int_\S \d \Th(\f, {\cal L}_\xi \f) \,-\, {\cal L}_\xi \Th(\f, \d \f)  \nn \\
&=& \oint_{\pl \S} \d Q_\xi \,-\, \in \Th \,.  \label{dH}
\eea
If one could find a $(n-1)$-form $B$ such that 
\be
\d \int_{\pl \S} \, \in B \;=\; \int_{\pl \S} \, \in \Th \;, \label{formB}
\ee
the Hamiltonian $H_\xi$ exists and is given by integrating (\ref{dH}): 
\be
H_\xi \;=\; \oint_{\pl \S} Q_\xi \,-\, \in B \;.
\ee
If we restrict $\S$ to be a subspace of a Cauchy surface in a diffeomorphic invariant way, $H_\xi$ could serve as a natural definition of the quasi-local energy\footnote{ Or {\it canonical energy}  according to Wald {\it et al}. We use the terminology ``quasi-local energy'' throughout the paper to refer to $H_\xi $ with respect to various vectors $\xi$\,. } 
 for the region $\S$ with respect to the vector $\xi$ \cite{Lashkari:2016idm}.

\hspace{5mm}

If $\xi$ generates a symmetry of the dynamical fields, i.e. ${\cal L}_\xi \f = 0$\,, it follows from ({\ref{HamiltonsEq}}) that $\d H_\xi = 0$\,, hence from (\ref{dH})
\be
\oint_{\pl \S} \d Q_\xi \,-\, \in \Th \;=\; 0  \,. \label{identity}
\ee
Consider a stationary black hole with bifurcate Killing horizon. Let $\,\z\,$ be the Killing vector field which generates the Killing horizon and vanishes on the bifurcation $(n-2)$-surface $\cal B$\,,  
\be
\z = \t + \OmH \psi \,, \label{zeta}
\ee
in which $\t$ and $\psi$ are the Killing fields generating the asymptotic time translation and the asymptotic rotation, respectively; $\,\OmH$ is the angular velocity of the horizon. Define the total energy $\cal E$ and the angular momentum $\cal J$ of the spacetime: 
\bea
{\cal E} &\equiv& \quad\, H_\t \;=\; \quad\, \oint_{\infty} Q_\t \,-\, {\mbox i}_\t B \;, \\
{\cal J} &\equiv& - \, H_\psi \;=\; - \, \oint_{\infty} Q_\psi \,-\, {\mbox i}_\psi B \; \
\eea
 (where an additional $-1$ factor has been introduced in the definition of the angular momentum to match the 
usual convention in general relativity). For hypersurface $\S$ with spatial infinity and $\cal B$ as its only boundaries, (\ref{identity}) takes the form
\be
\oint_{\cal B} \d Q_\z \,=\, \d {\cal E} \,-\, \OmH \,\d {\cal J} \,. \label{identity2}
\ee
$Q_\z$ depends on $\z$ only algebraically through $\z$ and $\na \z$\, since $\z$ is a Killing vector. Note that $\,\z |_{ \cal {\tiny B}} = 0\,$ and 
\be
\na_\m \z^\n |_{\cal B} \,=\, \kH n_\m^{~\,\n} \,, \label{nablazeta}
\ee
in which $\kH$ is the constant surface gravity of the Killing horizont and $n_{\m \n}$ is the binormal to $\cal B$ (normalized to $-2$)\,. Defining 
\be
S \,=\, {2 \p } \oint_{\cal B} {\hat Q}_\z \,, \label{Sdef}
\ee
where $ {\hat Q}_\z$ is obtained from $Q_\z$ by replacing $\na_\m \z_\n$ with $n_{\m \n}$\,, (\ref{identity2}) then gives rise to the first-law of black hole thermodynamics 
\be
\TH  \, \d S \;=\; \d {\cal E} \,-\, \OmH \,\d {\cal J} \,, 
\ee
in which $\TH = \kH / 2 \p$ is the Hawking temperature. It is concluded that the black hole entropy $S$ is proportional to the Noether charge associated with the horizon-generating Killing vector \cite{Wald:1993nt, Iyer:1994ys}. 

\hspace{5mm} 

The above approach is constructed for diffeomorphism invariant theory. For gravity theory with Chern-Simons term, the  action is diffeomorphism invariant only up to a surface term, i.e.,  
\be 
\d_\xi L \,=\, {\cal L}_\xi L \,+\, \dr \Xi_\xi \,.
\ee
In this case Wald's approach was generalized by Tachikawa \cite{Tachikawa:2006sz}. Both the 
Noether charge $\,Q_\xi\,$ and the Hamiltonian $\,H_\xi\,$ receive additional contributions coming from the surface term. By identifying the black hole entropy with the the horizon Noether charge plus additional modification terms, the first law is still satisfied. 

\hspace{5mm}

On the other hand, Wald's approach was constructed originally in the second order formalism of gravitational theory, in which the following fact is essential for the derivation of the black hole entropy:  the Noether charge $Q_\z$ depends algebraically on  $\na \z$, which is non-vanishing at the bifurcation surface $\cal B$. If one tries to generalize Wald's approach  straightforwardly to the first order formalism with vielbeins and spin connections, it seems that the black hole entropy would be vanishing, since  $Q_\z$ no longer depends on $\na \z$. This puzzle was clarified by Jacobson and Mohd \cite{Jacobson:2015uqa}. The crucial point is that in the first order formalism, $Q_\z$ depends on $\,\iz \om\,$, which is not vanishing at $\cal B$ as it looks at first glance. The reason is that  for vielbeins satisfying  
\be
\,{\cal L}_\xi e^a = 0 \,,
\ee
the  spin-connections $\om$ become divergent at the bifurcation surface, giving rise to finite $\iz \om\,$ hence non-vanishing black hole entropy. In general, however, the Lie derivative of a vielbein with respect to a Killing vector $\,\xi\,$ is non-zero, since the vielbein might undergo a Lorentz transformation under the flow generated by $\xi$\,. In this case, one need to introduce the so called Lorentz-Lie derivative 
\cite{Kosmann:1971, Jackiw:1979ub, Obukhov:2006ge, Fatibene:2011} instead, which contains a compensating local Lorentz transformation. The Lorentz-Lie derivative of $\,e^a$ with respect to a Killing vector are always vanishing, hence one could reformulate Wald's approach by replacing the Lie derivative with the Lorentz-Lie derivative, with the conclusion that the black hole entropy is the horizon Noether charge for a combination of diffeomorphism and local Lorentz symmetry \cite{Jacobson:2015uqa}.

\hspace{5mm} 

For Chern-Simons-like theories of gravity \cite{Bergshoeff:2014bia} in the first order formalism, the Lagrangian is invariant up to a surface term under the local Lorentz transformation, for which an approach incorporating the Lorentz-Lie derivative as well as the surface contribution was developed in \cite{Setare:2015nla, Setare:2015cvv, Setare:2015gss, Setare:2017wuj}, giving rise to a general formula for black hole entropy in such theories (see  \cite{Adami:2017phg} for a comprehensive review).

\hspace{2mm}


\subsection{The first order formalism revisited} \label{subsec:revisit}  

\hspace{5mm}

As mentioned in the above subsection, in the first order formalism, in general the Lie derivative of a vielbein with respect to a Killing vector field is non-zero, hence Wald's approach needs to be reformulated by introducing the Lorentz-Lie derivative. We notice, however, that for a large amount of geometries, it is not only possible but also natural to choose vielbeins that are invariant under the flow of some specific Killing vectors. In this case, a straightforward generalization of Wald's approach is possible. We are specially interested in gravity theories in three-dimension with the presence of torsion.

\hspace{5mm}

Considering a generic axially symmetric stationary spacetime in three dimension
\be \label{genericI}
\dr s^2 \;=\; - \, A(r)^2 N(r)^2 \dr t^2 \,+\, { \dr r^2 \over B(r)^2 N(r)^2}  
\,+\, C(r)^2 \left( \dr \f + N^\f (r) \dr t \right)^2\;,
\ee
it is natural to choose the ``diagonal'' vielbeins
\be \label{genericII}
e^0 \,=\, A(r) N(r) \dr t\,,  \quad~~ 
e^1 \,=\, {\dr r \over B(r) N(r)} \,, \quad~~
e^2 = C(r) \left( \dr \f + N^{\f}(r) \dr t \right) \,. 
\ee
Note that $e^a = e^a(r)$. For the Killing vector
\be \label{KillingV}
\,\xi = a\, \pl_t + b\, \pl_\f\,
\ee
 with $a,\,b$ constants, it follows immediately that 
\be
\cL_\xi e^a \,=\, 0 \,, \label{Liee}
\ee
since $\,e^a\,$ do not depend on $t,\,\f$. The Riemannian dual spin connections $\,\omt^a\,$, defined by torsion-free condition  $\,\dr e^a + \e^a_{~bc} \,\omt^b \,\w\, e^c = 0$\,, apparantly do not depend on $t,\,\f$, either. We further assume that the dual contorsion 1-form $ k^a$, defined by $\,T^a \,=\, \e^a_{~bc} \, k^b \,\w\, e^c \,$ with  $\,T^a \,$ the torsion 2-form, also do not depend on $t,\,\f$, which is natural due to the spacetime symmetry. The dual spin connection $\,\om^{a}\,$, defined by $\,T^a \,=\,\dr e^a + \e^a_{~bc} \,\om^b \,\w\, e^c $\, and take the form 
\be  
\om^a \,=\, \omt^a \,+\, k^a \; \label{om}
\ee
hence depend only on $r$ just like the vielbeins. It follows that the Lie-derivative of $\,\om^a\,$ with respect to $\,\xi\,$ also become vanishing 
\be
\cL_\xi \om^a \,=\, 0 \,. \label{Lieom}
\ee
Eq.(\ref{Liee})(\ref{Lieom}) ensure that Wald's origianl approach is still
valid in the first order formalism for diffeomorphism invariant gravity theories. In this case, there's no need to introduce the Lorentz-Lie derivative.

\hspace{5mm}

The above observation is particularly useful for gravity theories with Chern-Simons term. In the first order formalism, the Chern-Simons term is invariant under diffeomorphism (as manifestly shown in (\ref{actionMB4})), while is invariant only up to a surface term under the local Lorentz transformation. 
Our natural vielbeins (\ref{genericII}) and the dual spin connections (\ref{om}) are always invariant under the flow of $\,\xi$ (\ref{KillingV}), hence undergo no Lorentz transformation. As a result, there's simply no need to deal with the possible contribution arising from the surface term just as in procedures involving Lorentz-Lie derivatives.

\hspace{5mm}

The only possible problem, as pointed out in \cite{Jacobson:2015uqa} for Einstein gravity, is that the 
spin connections might become divergent at the the bifurcation surface, hence need to be treated carefully. Below we generalize the treatment for the spin connections in \cite{Jacobson:2015uqa} to gravity theories incorporating torsion. From (\ref{Liee}) we have 
\bea
0 \;=\; \cL_\xi e^a  &=&  \in \dr e^a \,+\, \dr \in e^a  \nn \\
&=& \in \Dr e^a \,+\, \Dr \in e^a \,-\, \left(\in \om^a_{~b} \right) \w\, e^b \;,  \label{Lie1}
\eea
in which $\Dr$ is the Lorentz covariant exterior derivative when acting on $p$-forms: 
\be
\Dr e^a \,=\, \dr e^a \,+\, \om^a_{~b} \,\w\, e^b \,=\, T^a\,. \label{D1}
\ee
Specifying the action of $\,\Dr$ on tensor 
indices to be the covariant derivative $\nd$\,, 
we interpret $\Dr$ to be the full derivative. The tetrad postulate states that
\be
\Dr_\m e^a_{~\n} \,=\, \pl_\m e^a_{~\n}  \,-\, \G^\l_{~\m\n} e^a_{~\l} 
\,+\, \om^a_{~b\m} e^b_{~\n} \,=\, 0 \,. \label{D2}
\ee
Substituting (\ref{D1})(\ref{D2}) into (\ref{Lie1}), we have
\bea
\in \om^a_{~b} 
&=& e^{~\m}_b \left(\in T^a\right)_\m \,+\, e^{~\m}_b e^{a}_{~\n} \nd_\m \xi^\n \,.
\eea 
Noticing that 
\bea
  {1 \over 2} \, \e^a_{~bc} \, e^{c\m} \,e^b_{~\n} \nd_\m \xi^\n   
&=& {1 \over 2} \, \e^a_{~bc} \, e^{c\m} \,e^b_{~\n}  \ndt_\m \xi^\n  
 \,-\,  {1 \over 2} \e^a_{~bc} \, e^{c\m} \,( \in T^b )_\m  \,+\, \in k^a   \;,
\eea
in which $\ndt$ is defined in terms of Levi-Civita connection, we get
\be \label{xi.omega}
\in \om^a \;=\; {1 \over 2} \, \e^a_{~bc} \, e^{c\m} \,e^b_{~\n}  \ndt_\m \xi^\n  
 \,+\, \in k^a   \;.
\ee
If we consider a Killing field $\,\z\,$ which generates a Killing horizon with surface gravity $\,\kH\,$, 
noticing (\ref{nablazeta}) we obtain
\be \label{zeta.omega}
\iz \om^a \,|_{\cal B} \;=\; - \, {1 \over 2}\, \kH \, \e^a_{~bc} \, n^{bc}
 \,+\, \iz k^a  \,|_{\cal B} \;.
\ee
Assuming that 
$\,\iz k^a\,$ is also finite on $\,\cB\,$, the limit of $\,\iz \om^a\,$ at the bifurcation surface turns out to have finite value. Again, the spin connection is divergent on $\,\cB\,$, which would be essential for giving rise to finite black hole entropy.

\hspace{5mm}

To summarize, for axially symmetric stationary spacetime and Killing fields of the form (\ref{KillingV}) in a diffeomorphism invariant gravitational theory, Wald's original approach could be applied straightforwardly in the first order formalism by just noticing the relation (\ref{zeta.omega}), without introducing the Lorentz-Lie derivative. As an explicit application, in the next section we study a specific three-dimensional gravity theory with torsion, i.e. the Mielke-Baekler gravity.

\hspace{2mm}


\section{Quasi-local energy in Mielke-Baekler gravity}  \label{sec:quasi}

\hspace{5mm}

In this section we investigate the quasi-local energy in the three-dimensional Mielke-Baekler (MB) gravity \cite{Mielke:1991nn,Baekler:1992ab}. This model was originally proposed as a torsional generalization of the topological massive gravity \cite{Deser:1981wh,Deser:1982vy,Deser:2002iw}, and could be formulated as a Poincar{\'e} gauge theory. The Lagrangian of the MB model contains the Einstein-Cartan term, the cosmological constant term, the Chern-Simons term for the curvature, as well as a translational Chern-Simons term linear in torsion and veilbein: 
\be \label{actionMB1}
 L \;=\; L_{\mbox{\tiny EC}}  \;+\;  L_{\L} \;+\; L_{\mbox{\tiny CS}}  \;+\; L_{\mbox{\tiny T}} 
 \;+\; L_{\mbox{\tiny M}}\;,
\ee
where
\bea 
L_{\mbox{\tiny EC}} &=& {1 \over \p} \, e^a \,\w\, R_a \;,  \label{actionMB2} \\ \nn \\
L_{\L}  &=& - \, { \L \over 6 \p} \, \e_{abc} \, e^a \,\w\, e^b \,\w\, e^c \;, \label{actionMB3} \\ \nn \\
L_{\mbox{\tiny CS}}  &=& -\, \tl \left( \om^a \,\w\, \dr \om_a 
\,+\, {1 \over 3} \,\e_{abc}\, \om^a \,\w\, \om^b \,\w\, \om^c \right)\;, \label{actionMB4} \\ \nn \\ 
L_{\mbox{\tiny T}}  &=& { \tt \over 2 \p^2} \, e^a \,\w\, T_a \;, \label{actionMB5}
\eea
in which $\L$ is the cosmological constant, $\tl,\;\tt$ are coupling constants, and $L_{\mbox{\tiny M}} $ is the Lagrangian of the matter. We have defined the dual curvature 2-form $\,R^{a} \,=\, {1 \over 2}\, \e^a_{~bc} R^{bc}\,$.
From now on we will focus on the case with vanishing source. 

\hspace{5mm}

Variation of the Lagrangian (\ref{actionMB1})-(\ref{actionMB5}) with respect to $e^a$ and $\om^a$ gives rise to the field equations: 
\bea
2 \p R_a \,+\,  2 \tt  T_a \,-\,  \p \L \, \e_{abc} \, e^b \,\w\, e^c &=& 0 \;,   \label{eom1}\\ 
2\p T_a \,-\, 4 \p^2 \tl \, R_a \,+\,  \tt  \,\e_{abc} \, e^b \,\w\, e^c &=& 0 \;.  \label{eom2}
\eea
Assuming that $ 1 + 2 \tt \tl  \neq 0\,$, the field equations are solved by 
\bea
T^a &=& { \cT \over \p } \,   \e^a_{~bc} \, e^b \,\w\, e^c \;, \label{T2} \\ \nn \\ 
R^a &=& - \, { \cR \over 2 \p^2 }   \,   \e^a_{~bc} \, e^b \,\w\, e^c \;, \label{R2}
\eea 
in which 
\be
\cT \,\equiv\, \frac{ -\tt \,+\, 2 \p^2 \L\tl  }{ 2 + 4 \tt \tl } \;,   \quad\quad 
\cR  \,\equiv\, - \, \frac{\tt^2 + \p^2 \L}{1 + 2 \tt \tl} \,. \label{calTR}
\ee
Apparently, eq.(\ref{T2}) is equivalent to 
\be
k^a \;=\; {\cT \over \p} \, e^a \;.  \label{k1}
\ee

\hspace{5mm}

The equations (\ref{T2})(\ref{R2})  have a family of BTZ-like solutions \cite{Garcia:2003nm}. The vielbeins take the same ``diagonal'' form as the BTZ black holes \cite{Banados:1992wn} 
\be
e^0 \,=\, N \dr t\,,  \quad~~ 
e^1 \,=\, {\dr r \over N} \,, \quad~~
e^2 = r \left( \dr \f + N^{\f} \dr t \right) \,, \label{eBTZ.1}
\ee
in which\footnote{The convention in \cite{Garcia:2003nm} is related to ours by the following replacement:  
$\chi \to 1$, the gravitational constant $\ell \to \p$, the effective cosmological constant $\Lef \to -\, \Lef\,$. }
\be \label{eBTZ.2}
N^2 (r) \,=\, -M - \Lef \, r^2 + {J^2 \over 4 r^2 } \,,  \quad\quad
N^\f (r) \,=\,  - \, { J \over 2 r^2 } \,
\ee
with
\be \label{eBTZ.3}
\Lef \,\equiv\, - \,\frac{\cT^2 + \cR }{ \p^2 } \;\,,
\ee
while the dual spin connections contain both the Riemannian part and the contorsion, according to eq.(\ref{k1})
\be \label{omBTZ.1}
\om^a \;=\; \omt^a \,+\, {\cT \over \p} \, e^a \,,
\ee
with the Riemannian part determined totally by the vielbeins (\ref{eBTZ.1}) through torsion-free condition
\be \label{omBTZ.2}
\omt^0 \,=\, N \dr \f\,, \quad~~ 
\omt^1 \,=\, - \, {N^\f \over N} \dr r \,,  \quad~~ 
\omt^2 \,=\, - \Lef \, r \dr t \,+\, r N^\f \dr \f  \,.
\ee
If $\tl = \tt = 0 $, we have $ \,\cT = 0 $ and $\Lef = \L$ , and the above solutions are reduced to the usual BTZ black holes.

\hspace{5mm}

The locations of the horizons are determined by $N^2(r) = 0$ :
\be
r_\pm^2 \;=\; {1 \over \, 2 \Lef} \left( - \, M \mp \sqrt{M^2 + \Lef J^2} \right) \,
\ee
(note that $\Lef < 0 $ for asymptotic AdS solutions). One could verify that 
\be \label{rprm}
r_+ \,r_- \;=\; { J \over 2 \sqrt{-\Lef}} \,. 
\ee
The angular velocity of the outer horizon is given by
\be
\OmH \, = \, -\, N^\f ( r_+) \,=\, {J \over 2 r_+^2} \,.
\ee
Vanishing of the conical singularity of the Euclidean BTZ metric gives the black hole temperature
\be
\TH \,=\, \frac{\Lef \left( r_+^2 - r_-^2 \right) }{2 \p r_+} \,, 
\ee
and the surface gravity is $\kH = 2 \p \TH$\,.

\hspace{2mm}


\subsection{Quasi-local energy} 

\hspace{5mm}

The Lagrangian (\ref{actionMB1})-(\ref{actionMB5}) is manifestly diffeomophism invariant. For spacetime with axially symmetry and Killing vectors of the form (\ref{KillingV}), we could derive the quasi-local energy based on our argument in Section {\ref{subsec:revisit}}. 

\hspace{5mm}

Variation of the Lagrangian (\ref{actionMB1})-(\ref{actionMB5}) with respect to the dynamical fields $e^a$ and $\om^a$ gives 
\be
\d L \;=\; \d e^a \,\w\, E_a^{(e)} \;+\; \d \om^a \,\w\, E_a^{(\om)} \;+\; \dr \Th (\f, \d \f) \,,
\ee
in which 
\bea
\Th (\f, \d \f) &=&{ 1 \over \pi} \,\d \om^a \,\w\, e_a 
\;+\; {\tt \over 2 \p^2} \, \d e^a \,\w\, e_a \;-\; \tl \, \d \om^a \,\w\, \om_a \,, \\ ~\nn \\
E_a^{(e)} &=& {1 \over \p} \left( R_a \,+\, {\tt \over \p} \, T_a 
\,-\, {\L \over 2} \, \e_{abc} \,e^b \,\w\, e^c \right) \;, \\
E_a^{(\om)}  &=&  {1 \over \p} \left( T_a \,-\, 2 \p \tl \, R_a 
\,+\, {\tt \over 2\p} \, \e_{abc} \, e^b \,\w\, e^c \right) \,.
\eea
$E_a^{(e)} = 0$ and $E_a^{(\om)} = 0$ give rise to the equations of motion (\ref{eom1})(\ref{eom2}). 
The diffeomorphism Noether current 3-form (\ref{NoetherJ}) turns out to be
\be
j_{\xi} \;=\; \dr Q_\xi \,+\, C_\xi \,,
\ee
in which 
\bea
Q_\xi &=& {1 \over \p} \, ( \in \om^a ) \,\w\, e_a \;+\; {\tt \over 2\p^2} \, (\in e^a) \,\w\, e_a 
\,-\, \tl \,(\in \om^a ) \,\w\, \om_a \,,  \\ ~\nn \\ 
C_\xi &=& -(\in e^a)  \,\w\, E_a^{(e)} \,-\, (\in \om^a )  \,\w\, E_a^{(\om)} \,.
\eea
Clearly the Noether current $\,j_\xi\,$ is exact on shell. 
The variation of quasi-local energy, according to (\ref{dH}), is given by 
\bea \label{deltaH}
\d H_\xi 
&=& \oint_{\,\pl \S} \left\{ \,
{ 1 \over \p } \left[\, (\in e^a) \,\w\, \d \om_a \;+\;  (\in \om^a) \,\w\, \d e_a \,\right]  \right. \nn \\ ~\nn \\
&& \quad\quad~ \left. \;+\; {\tt \over \p^2} \, (\in e^a) \,\w\, \d e_a \;-\; 2 \tl \,(\in \om^a) \,\w\, \d \om_a \,
\right\} \;.
\eea

To prove the integrability of the above equation, in general we need to set proper boundary condition on $\pl \S$. For the conserved charge and the black hole entropy in which we are interested, however, the integrability is straightforward, as can be seen below. 

\hspace{2mm}


\subsection{Mass and angular momentum} \label{sec:charge}

\hspace{5mm}

To obtain the variations of the total energy and the angular momentum of the BTZ-like solution (\ref{eBTZ.1})-(\ref{omBTZ.2}), we choose $\xi$ in expression (\ref{deltaH})  to be the Killing vector $\pl_t$ and $\pl_\f$,  respectively, and then integrate over the spatial infinity.  
By perturbing the parameters 
\be
M \,\to\, M \,+\, \d M\,, \quad \quad \quad J \,\to\, J \,+\, \d J \,,
\ee
the variations turn out to be
\bea
\d {\cal{E}} &\equiv& \quad \d H_{\pl_t} \, |_\infty \,\;=\; \d M \,-\, 2 \tl \left( {\cal{T}} \d M \,+\, \p \Lef \d J \right) \;, \label{dE} 
\\
\d {\cal{J}} &\equiv& - \, \d H_{\pl_\f} \, |_\infty \,\;=\; \d J ~ \,+\, 2 \tl \left( \p \d M \,-\, {\cal{T}} \d J \right) \,. \label{dJ}
\eea
Obviously the above expressions are integrable. Assuming that the pure AdS$_3$ is the ground state, i.e., 
${\cal{E}} = {\cal{J}} = 0$ for $M = J = 0$\,, the total energy and the angular momentum are obtained by integrating (\ref{dE})(\ref{dJ}):
\bea
{\cal{E}}  &=&  M \,-\, 2 \tl \left( {\cal{T}}  M \,+\, \p \Lef J \right) \;,  \label{energy} \\
 {\cal{J}} &=&   J ~ \,+\, 2 \tl \left( \p  M \,-\, {\cal{T}}  J \right) \,, \label{angularmomentum}
\eea
which are consistent with the previous results in the literature obtained via different approaches \cite{Garcia:2003nm, Blagojevic:2006jk}\,.

\hspace{2mm}


\subsection{Black hole entropy} \label{sec:entropy}

\hspace{5mm}

To derive the black hole entropy, we choose $\xi$ in expression (\ref{deltaH})  to be
\be
\z = \pl_t \,+\, \OmH \pl_\f\,,
\ee
and integrate over the bifurcation surface $\,{\cal B}\,$ at $\,r \,=\, r_+\,\,$. 
Note that for a generic axially symmetric stationary black hole of the form (\ref{genericII}), the location of the horizon $\,r_+\,$ as well as the angular velocity of the horizon $\,\OmH\,$ are determined by 
\be N(r_+) = 0\,, \quad\quad\quad \OmH = -\,N^\f (r_+) \,, \ee separately, from which it is straightforward to check that
\be \label{xi.e}
\iz e^a \,|_{{\cal B}} \;=\; 0 \;.
\ee
Since the dual contorsion 1-form $k^a$ is proportional to $e^a$ according to the equation of motion (\ref{k1}), 
we also have
\be \label{xi.k}
\iz k^a \,|_{{\cal B}} \;=\; 0 \,.
\ee
As analyzed in Subsection \ref{subsec:revisit}, the vanishing of the Lie derivative of $e^a$ with respect to Killing vector 
$\z$ indicates (\ref{zeta.omega}). Noticing (\ref{xi.k}), we have
\be
\iz \om^a \,|_{{\cal B}} \;=\; - \,{1 \over 2}\,\kH \,\e^a_{~bc} \,n^{bc} \,.
\ee
For simplicity, we introduce
\be
N^a \;\equiv\;  {1 \over 2}\,\e^a_{~bc} \,n^{bc}\,,
\ee
hence
\be
\iz \om^a \,|_{{\cal B}} \;=\; -\, \kH N^a \,. \label{izw}
\ee
Noticing (\ref{xi.e})(\ref{izw}), the variation of the quasi-local energy (\ref{deltaH}) integrated on the bifurcation surface $\cal B$ with respect to $\,\z\,$ turns out to be
\bea
\d H_\zeta 
&=&  -\, \kH \oint_{\,{\cal B}} N^a \left( {1 \over \p} \d e_a \,-\, 2 \tl \d \om_a \right) \,. 
\eea
Since $\,N^a$ is just a constant vector on $\cal B$, the above expression is obviously integrable and gives rise to
\be
H_\zeta 
\;=\; -\, \kH \oint_{\,{\cal B}} N^a \left( {1 \over \p}  e_a \,-\, 2 \tl  \om_a \right)
\ee
Since the only non-zero components of the binormal to the bifurcation surface $\cal B$ is $n_{01} = -\, n_{10} = 1$,
the only non-zero component of $N^a$ is just $N^2 = -1$\,.
Motivated by (\ref{Sdef}), we define the black hole entropy by replacing $H_\z$ with $H_\z / \kH$ and multiplying a factor $2 \pi$, obtaining a general formula for the entropy of rotating black holes in the MB model: 
\be \label{entropyformula}
S \;=\; {2 \p \over \kH} \, H_\zeta  \;=\; 2 \p \oint_{\,\cal B}  \left( {1 \over \p}  e^2 \,-\, 2 \tl  \om^2 \right)\,,
\ee
which contains explicit contribution coming from the Chern-Simons action. The first law of black hole thermodynamics 
\be
\TH \d S \;=\; \d {\cal M} \;-\; \OmH \d {\cal J}
\ee
is satisfied by construction.

\hspace{5mm}

By submitting explicitly the solution (\ref{eBTZ.1})-(\ref{omBTZ.2}) into our entropy formula (\ref{entropyformula}) and noticing (\ref{rprm}), we obtain the black hole entropy for the BTZ-like solutions in the MB model: 
\be \label{entropy}
S \;=\; 4\p r_+ \,-\, 8 \p \tl \left( {\cal T} r_+ \,-\, { \p \over \,\lef\, }\,r_- \right)\;,
\ee
in which we have introduced an effective AdS radius $\lef^{-2} \equiv - \Lef$\,. This is 
consistent with previous result in \cite{Blagojevic:2006jk} via the Euclidean partition function approach\footnote{The convention in \cite{Blagojevic:2006jk}  is related to ours by replacing $\,G \to 1 / 8\,,~ \a_3 \to - \tl\,, ~p \to 2 {\cal T} / \p\,$ and $\,\ell \to \lef$\,.},  as well as that 
in \cite{Ma:2013eaa} via a direct integration from the first law. Our entropy formula (\ref{entropyformula}) is also consistent with the generic expression for black hole entropy in Chern-Simons-like theories in Section XVI of \cite{Adami:2017phg} via a different approach employing the Lorentz-Lie derivative.

\hspace{5mm}

Before conclusion, we give a brief discussion on the torsional effect. The torsion of the BTZ-like solution (\ref{eBTZ.1})-(\ref{omBTZ.2}) is characterized by the parameter $\cal{T}$, which originates from the non-vanishing coupling $\tt\,,~\tl\,$, according to (\ref{calTR}). It is explicit that the total energy (\ref{energy}),  the angular momentum (\ref{angularmomentum}) as well as the black hole entropy (\ref{entropy}) contain explicit contribution from the  torsion. On the other hand, for solutions with vanishing torsion, i.e., ${\cal T} = 0$\,, the conserved charges as well as the black hole entropy still differ from those in Einstein gravity, due to the Chern-Simons coupling $\tl$\,. In general, the quasi-local energy (\ref{deltaH}) receives corrections from both the $\tt$-term and the $\tl$-term, however, it turns out that only the  $\tl$-term gives non-vanishing contribution to the total energy, the angular momentum as well as the black hole entropy.

\hspace{2mm}


\section{Conclusion and discussion} \label{sec:conclusion}

\hspace{5mm}

In this paper, we showed that for axially symmetric stationary spacetime and specific Killing vectors, Wald's approach could be applied straightforwardly in the first order formalism. Using this method, we studied the quasi-local energy in three dimensional torsional Mielke-Baekler gravity, derived the general entropy formula for stationary black holes, reproduced precisely the total energy, the angular momentum as well as the black hole entropy for the BTZ-like solutions. 

\hspace{5mm}

It should be pointed out, however, that if one tries to consider more general Killing vectors other than (\ref{KillingV}), the Lie derivative of the vielbeins might not be vanishing, in which cases the Lorentz-Lie derivative would need to be introduced. One example is the Killing vector that vanishes on the boundary of the entanglement wedge in AdS spacetime. It has been demonstrated in \cite{Lashkari:2015hha, Lashkari:2016idm} that the quasi-local energy in such entanglement wedge in Einstein gravity is equivalent to the so-called relative entropy in the dual boundary CFT. This equivalence was generalized to the Einstein-Cartan gravity in \cite{Lin:2016fua,Ko:2017iki}. It would be interesting to investigate further if there are similar stories for three dimensional torsional MB gravity \cite{Ning:2017x}.

\hspace{2mm}

\subsection*{Acknowledgements}

We would like to thank Feng-Li Lin and Haitang Yang for careful reading of the manuscript and helpful comments and suggestions. BN also thanks the participants of the advanced workshop ``Dark Energy and Fundamental Theory" supported by the Special Fund for Theoretical Physics from the National Natural Science Foundations of China with Grant No.~11447613 for stimulating discussion. This work is supported by the National Natural Science Foundation of China with Grant No.~11505119 and 11975158.

\hspace{2mm}


\providecommand{\href}[2]{#2}\begingroup\raggedright\endgroup

\end{document}